\documentclass[12pt,preprint]{aastex}
\shorttitle{M31 Globular Cluster Velocity Dispersions}
\shortauthors{Strader et al.}
\def\etal{{\it et al.}}
\usepackage{subfigure}

\begin{document}

\title{Mass-to-Light Ratios for M31 Globular Clusters: Age-Dating and a Surprising Metallicity Trend}

\author{Jay Strader\altaffilmark{1,2}, Graeme H. Smith\altaffilmark{3}, Soeren Larsen\altaffilmark{4}, Jean P. Brodie\altaffilmark{3}, John P. Huchra\altaffilmark{1}}
\email{jstrader@cfa.harvard.edu}

\altaffiltext{1}{Harvard-Smithsonian Center for Astrophysics, Cambridge, MA 02138}
\altaffiltext{2}{Hubble Fellow}
\altaffiltext{3}{UCO/Lick Observatory, Santa Cruz, CA, 95064}
\altaffiltext{4}{Utrecht University, Utrecht, Netherlands}

\begin{abstract}

We have obtained velocity dispersions from Keck high-resolution integrated 
spectroscopy of ten M31 globular clusters (GCs), including three candidate 
intermediate-age GCs. We show that these candidates have the same $V$-band 
mass-to-light ($M/L_V$) ratios as the other GCs, implying that they are likely 
to be old. We also find a trend of 
derived velocity dispersion with wavelength, but cannot distinguish between a 
systematic error and a physical effect.

Our new measurements are combined with photometric and spectroscopic data from 
the literature in a reanalysis of all M31 GC $M/L_V$ values. In a combined
sample of 27 GCs, we show that the 
metal-rich GCs have \emph{lower} $M/L_V$ than the metal-poor GCs, in conflict 
with predictions from stellar population models. Fragmentary data for other 
galaxies support this observation.

The M31 GC fundamental plane is extremely tight, and we follow up an earlier 
suggestion by Djorgovski to show that the fundamental plane can be used to 
estimate accurate distances (potentially 10\% or better).

\end{abstract}


\section{Introduction}

A consistent theme in the study of globular clusters (GCs) in M31, going back to the 
earliest works in the subject, has been the identification of alleged differences between 
the GC systems of M31 and the Milky Way. These include GC luminosity (Hubble 1932), mean 
metallicity (van den Bergh 1969), and abundance ratios (Burstein \etal~1984). In most 
cases, the discrepancies are reduced or disappear in subsequent work (see the review of 
Brodie \& Strader 2006 and references therein).

Arguably, the central topic of M31 GC work in this decade is cluster ages. A number of 
recent papers (Beasley \etal~2005; Burstein \etal~2004; Fusi Pecci \etal~2005; Puzia, 
Perrett, \& Bridges 2005) have presented evidence, mostly from low-resolution 
spectroscopy, that M31 has a significant population of intermediate-age (1--6 Gyr) 
globular clusters (GCs). The Milky Way has no known such clusters, and these results lead 
one to wonder whether the GC formation history (and thus perhaps the violent star 
formation history) of M31 was very different than the Milky Way. Such an interpretation 
could be consistent with the extensive accretion history of M31, as indicated by the 
presence of large stellar streams and plumes in the halo (e.g., Ibata \etal~2001).

The standard approach to deriving GC ages from low-resolution spectroscopy is to use a 
combination of Balmer and metal lines to break the age-metallicity degeneracy inherent in 
any single line. At fixed metallicity, stronger Balmer lines indicate a hotter main 
sequence turnoff and younger ages. However, in old GCs a variety of hot star 
populations---most notably blue horizontal branch stars---can lead to ``artificially" 
strong Balmer lines and thus to an underestimation of GC ages.

There are several ways to address this concern, including color-magnitude diagrams 
reaching below the horizontal branch and the direct detection of the integrated light of 
such stars using UV photometry. A different approach was that taken by Larsen \etal~(2002) 
in studying putative intermediate-age GCs in M33. Mass-to-light ($M/L$) ratios are 
derived from high-resolution integrated spectroscopy and Hubble Space Telescope (HST) 
imaging. The spectra yield velocity dispersions, and when combined with a radius measured 
from the HST imaging, an application of the virial theorem gives the GC mass. Stellar 
population models generically predict that $V$-band $M/L$ ($M/L_V$) increases with age; 
as a specific example, Maraston (2005) models using a Kroupa initial mass function (IMF) 
indicate that ($M/L_V$) increases by approximately a factor of two as the clusters age 
from 5 to 12 Gyr at fixed metallicity. Younger clusters are predicted to have an even larger difference. 
This factor of two is easily detectable with the $\sim$ 15\% -- 20\% relative errors in 
$M/L_{V}$ that can be achieved with high-quality spectroscopy.

As part of a larger program to obtain $M/L_V$ ratios from high-resolution spectra of M31 
GCs, we have obtained observations of three objects posited to have intermediate ages on 
the basis of low-resolution data. In the first part of this paper we compare derived 
$M/L_V$ values for these three GCs to measurements for a larger sample of probable old 
M31 GCs, both from our own observations and the literature.

When coupled with new structural data from HST, the number of $M/L_V$ measurements now 
available for M31 GCs is large enough to warrant a reanalysis of their dynamical 
parameters. Such an effort constitutes the second part of this paper, which focuses on an 
unexpected anti-correlation between $M/L_{V}$ and cluster metallicity and on the joint 
Milky Way--M31 GC fundamental plane.

\section{Data}

We have acquired high resolution spectra of a sample of M31 GCs, taken over a ten-year 
period for a variety of programs, including a stellar population study. Three of the 
clusters observed were chosen because of a suspected intermediate age; the remainder were 
generally selected to have high luminosities and to span a range of metallicities. The 
signal-to-noise (S/N) of these observations is larger than necessary to derive velocity 
dispersions; the resulting dynamical masses are among the most precise obtained for any 
cluster outside the Galaxy. In addition to the M31 GCs in the sample, we include a single 
object in M33 that is also a candidate intermediate-age GC.

The three candidate intermediate-age GCs were selected from Burstein \etal~(2004) and 
Beasley \etal~(2005), with the following age estimates: 126-184 (3--6 Gyr), 232-286 (2--5 
Gyr), and 311-033 (5 Gyr). Puzia \etal~(2005) argue that these GCs are all older than 6 Gyr, 
indicating the general difficulty of obtaining consistent age estimates from optical 
spectroscopy.

Observations were made with Keck/HIRES between 1997 and 2007. Table 1 is an observing log, 
indicating the exposure times, slit sizes, and dates of the observations. The Barmby 
\etal~(2000) cluster naming convention is used. All runs used the red-optimized 
cross-disperser and collimator. The 1997 data used the A11 decker with a width of 
$0.87\arcsec$ and a length of $5\arcsec$. In 2004 HIRES was upgraded to a three-chip mosaic 
with smaller pixels and wider wavelength coverage, and all observations from 2005 onward 
used the C1 decker ($0.86\arcsec \times 7\arcsec$; $R \sim 45000$) binned spatially by 2 
pixels. The full-width at half-maximum (FWHM) of all of the spectra is $\sim 6.5-7$ km/s.

Spectra taken from 2005-2007, after the HIRES upgrade, were reduced using the IDL HIRedux 
package\footnote{http://www.ucolick.org/\textasciitilde xavier/HIRedux/}. Images were 
bias-corrected and flat-fielded. A conservative cosmic ray rejection was used on the 2-D 
images for clusters with multiple exposures. The object spectra were then traced, 
sky-subtracted, and optimally extracted order by order. The spectra were wavelength 
calibrated with a series of ThAr arcs taken throughout the night. The 1997 data were reduced 
with MAKEE\footnote{http://spider.ipac.caltech.edu/staff/tab/makee/}; see Larsen 
\etal~(2002) for details of the data reduction.

Template stars, ranging in spectral type from late F to K, were observed on each run. 
Typically 10-15 stars were observed per run, including both dwarfs and giants. These stars 
were reduced in a very similar manner to the GC data, except that sky subtraction was 
generally not necessary before extraction.

\section{Analysis}

\subsection{Velocity Dispersions}

To estimate velocity dispersions ($\sigma$) through cross-correlation, one must determine 
the relationship between the FWHM of the cross-correlation peak and the velocity dispersion. 
This was done by convolving a template star with Gaussians in steps of 1 km/s, then 
cross-correlating the convolved templates with the original star (Figure 1 shows an example 
of the FWHM--$\sigma$ relationship). A spline fit was used for interpolation between the 1 
km/s steps. In general these fits are similar for different template stars but not exactly 
so, owing to the different temperatures, gravities, and abundances of the stars. No template 
star matches GC spectra perfectly, which can be roughly understood as a combined G dwarf and 
K giant spectrum in a proportion that varies with wavelength.

We follow the method of Larsen \etal~(2002) in improving the robustness of the procedure by 
cross-correlating each template with each other convolved template. For $N$ total template 
stars, there are $N^2$ different relationships between the velocity dispersion and the FWHM of 
the cross-correlation peak, and $N^2$ estimates of the velocity dispersion for each order. While 
we cannot assume that these individual estimates capture all of the uncertainty in the process 
of deriving velocity dispersions, the root mean square (rms) dispersion of these values does 
give a sense of the relative uncertainty in the velocity dispersion estimate for each order. 
There were no strong differences in the velocity dispersions estimated from different templates 
(see Figure 2, described below), though some stars, for example the most metal-poor ones, had 
lower cross-correlation amplitudes and correspondingly larger errors in the velocity dispersion. 
The typical rms dispersion among template pairs was $\sim 0.4-0.5$ km/s per order, with most 
values below 1 km/s.

Once these relationships were estimated, GC velocity dispersions were derived through 
order-by-order cross-correlation with template stars. The wavelength range and number of 
echelle orders varied with the exact setup and whether the data were taken before or after 
the HIRES upgrade. For the pre-upgrade spectra, the typical wavelength coverage was $\sim 
3730-6170$ \AA, covering 38 orders. The post-upgrade spectra cover $\sim 3840-8360$ \AA\ 
over 51 orders. Not all orders were used for the final velocity dispersion estimates. The 
bluest orders typically had low S/N, and many of the red orders were contaminated by 
telluric features or by poor sky subtraction. Also excluded were some orders with lines with 
strong damping wings (e.g., Balmer lines), since these wings are often not well-reproduced 
in the template stars and thus lead to poor estimates of the velocity dispersion. The number 
of orders used in the final value varied among the clusters but was typically 20-30.

We experimented with several methods for combining the projected velocity dispersion 
($\sigma_p$) from different orders into a composite value, including averages weighted by the 
mean cross-correlation peak height or by the rms scatter among the template fits. The final 
$\sigma_p$ listed in Table 2 are Tukey biweights of the values from the individual orders, with 
the error taken as the rms among the orders. These estimates closely match those from the other 
averaging methods in nearly all cases, but are less sensitive to occasional outlying orders.  
Figure 2 shows the order-by-order values of $\sigma_p$ for 126-184, with the rms variation among 
different template pairs plotted as error bars. The shaded region represents the 1$\sigma$ 
boundary of the overall $\sigma_p$ estimate. This figure suggests that the total error budget of 
$\sigma_p$ is dominated by systematic variation among orders rather than by variation among 
templates.

Four GCs in our sample were observed twice; the first three in 1997 and 2007, and the last 
pair in 2005 and 2007. A comparison of the estimates, with the earlier value listed first, 
shows astonishing internal agreement: 163-217 (18.2 vs.~18.3 km/s), 171-222 (15.0 vs.~15.0 
km/s), 225-280 (28.4 vs.~28.5 km/s), and 358-219 (11.8 vs.~12.1 km/s). For these clusters we 
adopt the straight average as the final estimate. These results give us confidence in the 
relative homogeneity of our measurements, even though they were taken over a very wide time 
baseline. These observations attest to the excellent long-term stability of HIRES, even 
spanning its major upgrade.

The M31 GC with the lowest estimated dispersion is 126-184, with $\sigma_p = 7.2\pm0.9$. 
While this value of $\sigma_p$ is low and comparable to our resolution, an inspection of 
Figure 1 shows that it is still on the linear part of the curve relating $\sigma_p$ to the 
measured FWHM from cross-correlation. Thus, there should be no bias in the estimate of 
$\sigma_p$ for this cluster, though the relative error is larger than for the other M31 
GCs.

Table 2 gives the observed velocity dispersions $\sigma_p$ of our sample. For two of the 
objects, 034-096 and M33 GC U77, there is no high-resolution imaging available, and so we 
cannot derive dynamical masses at the present time. Table 2 also gives heliocentric radial 
velocities for GCs with data taken from 2005 onward. While many template stars were 
observed on each run, most of these are not radial velocity standards, and systematic 
offsets on the order of 1--2 km/s cannot be ruled out.

\subsection{Trends with Wavelength}

A curious feature of the spectra taken in 2007 is that there is a weak trend of $\sigma_p$ 
with wavelength. Figure 3 shows the metal-rich GC 163-217 as an example. $\sigma_p$ 
decreases essentially monotonically from the blue to the red, dropping by $\sim 1$ km/s over 
$\sim 2000$ \AA\ in wavelength. Four of the six GCs from the 2007 run show a similar trend, 
with the amplitude varying from $0.3-1.5$ km/s over the full range of the spectra (the 
order-to-order scatter in the 1997 data is much larger and no trend is visible).  The other 
two objects have data of equal quality but show no trend. The relationship is not dependent 
on a single type of template star---it persists even if single templates of varying 
temperatures or gravities are used. In particular, our set of templates spans late-G to 
mid-K subgiants and giants and late-F to late-G dwarfs. That this effect is not uniformly 
observed in all GCs also suggests that it is not due to an instrumental systematic, e.g., 
focus variations somehow not properly captured by the use of template stars.

To our knowledge no similar effect has ever been observed in an extragalactic GC. This is 
probably due to the extremely high S/N of these 2007 data (50-80 per resolution element).

This trend could be either a systematic issue with the derivation of $\sigma_p$ or a real 
physical effect. In the former case, the likely cause would be an increasing contribution of 
dwarf light at blue wavelengths, which could conceivably give higher $\sigma_p$ in a 
template mismatch with giants. As mentioned above, the trend persists even if a dwarf 
template is used, but perhaps the problem is more complex; light from blue stragglers and 
horizontal branch stars may also make an important contribution between 4000 and 4500 \AA. 
This could be tested by doing a stellar population synthesis with the appropriate template 
stars, and using the resulting spectrum as a template instead of a single star. 
Unfortunately, we have not yet observed a sufficient range of stars to try this 
synthesis properly.

As a first step, we took a G dwarf and K giant, normalized the spectra, and coadded them 
with wavelength-dependent weights appropriate for an old metal-rich stellar population 
(like 163-217). This combined spectrum was then used to derive $\sigma_p$ order-by-order in 
the same way as for single templates. The trend apparent in Figure 3 persisted, indicating 
that it is unlikely to be due solely to a changing mix of turnoff dwarfs and red giants.

If the trend is not due to a systematic error, we suggest that a possible cause is mass 
segregation. Mass segregation is well-established among main sequence stars in most 
Galactic GCs (Meylan \& Heggie 1997), although the color gradients observed in some 
GCs, especially in UV colors, are more likely due to post-main sequence stars such as 
blue stragglers and blue horizontal branch stars (Djorgovski \& Piotto 1993). Blue 
stragglers, in particular, are preferentially found in the cores of GCs (Dalessandro 
\etal~2008 and references therein), and could substantially contribute to the cluster 
light in the blue. It is worth recognizing that Barmby \etal~(2007; hereafter B07) 
found no significant differences in the core or half-light radii or total integrated 
model luminosities between $V$ and $I$-equivalent imaging of the same M31 GCs, but did 
observe systematic offsets in the central surface brightnesses between the two bands. 
This may indicate the existence of color gradients in luminous M31 GCs, but this 
conclusion is not secure because of uncertainties about the effects of saturation in 
the central regions in the HST imaging.

A first-order analysis of mass segregation can be done as follows. Assume that there are 
two populations of stars: red giants that dominate in the red and turnoff dwarfs that 
dominate in the blue. For an old metal-rich cluster the dwarfs will have masses of 
0.9--0.95 $M_{\odot}$, while the giants will be slightly more massive---around 1 
$M_{\odot}$ before they start to lose significant mass on the red giant branch (e.g., 
Marigo \etal~2008). If the two populations are thermalized then the condition 
$\sigma_{giant}^2/\sigma_{dwarf}^2 \sim m_{dwarf}/m_{giant}$ should hold. For the listed 
masses this corresponds to $\sigma_{giant}^2/\sigma_{dwarf}^2 \sim 0.9-0.95$. If we use 
163-217 (the cluster plotted in Figure 3) as an example, the observed ratio is 
$\sigma_{giant}^2/\sigma_{dwarf}^2 \sim 0.9$, within the range of the predicted effect.

\subsection{Masses}

As discussed in Larsen \etal~(2002) and Dubath \& Grillmair (1997; DG97), there are two 
standard methods for estimating GC masses. The first is the virial theorem:

\begin{equation}
M_{vir} = \frac{7.5 \sigma_{\infty}^{2} r_{hm}}{G} 
\end{equation}

\noindent 
where $\sigma_{\infty}$ is the global velocity dispersion and $r_{hm}$ is the half-mass radius. $r_{hm}$ is generally 
$\sim 4/3$ of the half-light radius $r_h$, but for GCs fit by King models, it can be obtained directly 
by integration.

The second common method is a King mass, given by:

\begin{equation}
M_{king} = a\frac{\sigma_{0}^{2} r_{c}}{G}
\end{equation}

\noindent 
using the core radius $r_c$, the central velocity dispersion $\sigma_0$, and the constant 
$a$, which is a function of the cluster structure.

At the distance of M31, where GCs are marginally resolved from the ground, the observed 
projected velocity dispersions $\sigma_p$ are not accurate estimates of either the central 
or global velocity dispersion ($\sigma_0$ and $\sigma_{\infty}$). Previous works generally 
used representative corrections of $\sigma_p$ to $\sigma_0$ and $\sigma_{\infty}$, which are 
of order 10\% for typical slit widths used to observe M31 GCs. The sense of the corrections 
is that $\sigma_0$ is higher than $\sigma_p$ and $\sigma_{\infty}$ lower.

B07 provides homogeneous King model fits for nearly all of the GCs in our sample (the 
exceptions are discussed in \S 3.6.1), allowing us to derive aperture corrections on a 
cluster-by-cluster basis. This was done by integrating the published King model over the 
slit width and length used for each cluster. The values of $\sigma_0$ and $\sigma_{\infty}$ 
thus obtained are listed in Table 3. 

Since the quantities in both $M_{vir}$ and $M_{king}$ are based on the same King model fits, 
these mass estimates are not independent; in fact, their ratio depends only on the cluster 
concentration (Larsen \etal~2002). We list both estimates in Table 3 but adopt the virial 
masses for the remainder of the paper.

The errors in the masses were obtained from standard propagation of errors in the radii and 
$\sigma_p$; errors in the latter dominate, as expected when high-quality HST imaging is 
used. The binding energy ($E_b$) as defined by McLaughlin (2000) is also listed in Table 3; 
the formal errors on this quantity are generally large (20-100\%) and are not given. 
The 
error introduced in $\sigma_p$ by the uncertainty of centering the cluster in the slit is 
negligible (offset integration of a range of King models gives a difference of
$\la 0.1$ km/s for an offset of 0.3\arcsec).

\subsection{Comparison with Previous Works}

A number of our GCs have either repeat velocity dispersion estimates from our own observing 
runs or published values already in the literature. The two relevant modern studies are 
Djorgovski \etal~(1997; D97) and DG97, who presented data for 21 and 10 GCs respectively.

D97 used Keck/HIRES with a $1.15\arcsec \times 7.0 \arcsec$ slit and DG97 used Lick/Hamspec 
with two different slit widths (1.5\arcsec and $2.0\arcsec \times 5\arcsec$). Both of these 
are wider than our slit ($0.86\arcsec \times 7.0\arcsec$) and so the native velocity 
dispersion measurements cannot be directly compared. Instead we can compare the central 
velocity dispersions ($\sigma_0$) derived through integration of King models in the same way 
as with our data.

These comparisons can be found in Table 4 for the three GCs in common among the studies. For 
two of the GCs, the agreement is good. However, our value of $\sigma_0$ for 358-219 (13.0 
km/s) is $\sim 2\sigma$ larger than either of the other published estimates.  We have two 
separate observations of this GC, one from 2005 and one from 2007 (the latter of which has 
very high S/N), which differ by only 0.3 km/s. Using the low $\sigma_0$ values of D97 and 
DG97 would give an unusually low mass-to-light ratio for this apparently normal old cluster 
(Puzia \etal~2005 give a formal spectroscopic age of 10 Gyr and find [Fe/H] $\sim -2$). The 
very low metallicity of the cluster, together with the faint central surface brightness, may 
have led to a paucity of strong lines and lower S/N than GCs of similar luminosity, 
complicating the estimation of $\sigma_p$ in previous works.

\subsection{Homogenizing the Data}

The availability of homogeneous structural parameters from B07 for most of the M31 GCs with 
dynamical data motivates us to rederive dynamical quantities for GCs from previous works. 
This was done identically to the analysis for our spectra. Using the published integrated 
velocity dispersion and reported apertures, we integrated over King models to derive 
$\sigma_0$, $\sigma_{\infty}$, King and virial masses, and $E_b$. For the five GCs in common 
between D97 and DG97, we simply report straight averages of the derived quantities. For each 
of these clusters the values of $\sigma_0$ were consistent within 1$\sigma$, and there was 
no systematic trend between the studies. Cluster 000-001 (G1) was observed by both D97 and 
Cohen (2006), and the estimates are less consistent ($28.1 \pm 1.9$ and $24.2 \pm 
1.7$, respectively), but we again adopted the average as the final value. The three GCs in 
common with our sample are omitted, as are two GCs with $\sigma_p$ from D97 or DG97 without 
published high-resolution structural data (193-244 and 218-272). Also omitted is 147-199 
from DG97. On the basis of its very low $\sigma_p$ (formally $\sim 2.4$ km/s using their 
Equation 1) and radial velocity, they argued that it was a star; HST imaging from B07 shows 
that it is clearly a GC. In fact, the radial velocity of $-51$ km/s, while not near the 
systemic velocity of M31, is well within the range typical of M31 GCs (e.g., Lee 
\etal~2008), but we cannot consider the reported $\sigma_p$ as reliable.

These derived quantities are all given in Table 5. The published values of $\sigma_p$ are 
given in both B07 and in the original papers and we do not reproduce them here. Table 5 also 
contains the total $V$-band luminosities of the GCs from B07 and the corresponding King and 
virial $M/L_V$. The derivation of cluster luminosities is discussed below.

Tables 3 and 5 together summarize the current knowledge of integrated dynamical information 
for M31 GCs.

\subsection{Mass-to-Light Ratios}

\subsubsection{New Structural Parameters}

As discussed in the Introduction, a principal goal of this paper is to compare $M/L_V$ for 
candidate intermediate-age GCs with old GCs. Most of the GCs in our sample (listed in Table 
3) have structural parameters from B07, but several do not. However, these clusters 
(126-184, 163-217, 171-222) do have archival HST imaging from WFPC2 or ACS. The first two 
objects have WFPC2/F555W data, but at the time of the analysis, only ACS/F435W data was 
available for 171-222. As the reddening for this cluster is well-constrained we are not 
overly concerned with its inclusion in the sample.

We derived core and half-light radii, concentration, and central surface brightnesses 
through fitting King models, as described in detail in Larsen \etal~(2002). These parameters 
are listed in Table 6. To ensure that there is no systematic difference between our 
structural fits and those of B07, we also fit King models for those objects also in B07 
using the same imaging. The rms difference in $r_c$ was 0.03 pc, indicating good agreement 
between the fitting procedures.

\subsubsection{Cluster Luminosities}

For GCs with structural data from B07, we use their integrated, extinction-corrected 
$V$-band model luminosities. See B07 for an extensive description of the photometry and 
comparison with ground-based measurements. These luminosities assume a constant distance of 
780 kpc. The luminosity errors listed in Tables 3 and 5 are the formal errors from B07, but 
for all but one cluster we have added additional errors of 5\% to the $M/L_V$ values to 
account for uncertainties in distance and reddening.

The highly reddened cluster 037-327 has been discussed extensively in the literature 
(Barmby \etal~2002; Ma \etal~2006; Cohen 2006; B07) and deserves special mention. 
The published value of $E(B-V)$ is $\sim 1.3$ with a large error, principally due to 
differential reddening. In \S 4 we estimate the reddening needed to place 037-327 on 
the GC fundamental plane of core parameters and find $E(B-V) = 0.92$. We adopt this 
reddening for the remainder of the paper, but we add an additional error of 75\% to the 
$M/L_V$ values, corresponding roughly to an uncertainty of 0.2--0.25 in $E(B-V)$, since 
the central value of the extinction may not be appropriate for the entire cluster (for 
the same reason, its $(V-K)_0$ color is also very uncertain).

For the three GCs with new structural parameters, we use integrated $V$ magnitudes from 
Caldwell \etal~(2009) and reddenings from Barmby \etal~(2000) to estimate luminosities. 
These clusters are marked with an asterisk in Table 3, and the listed errors include 
the uncertainty in the reddening.

\subsubsection{$M/L_V$ and Analysis}

Combining our new measurements with those from the literature, we have 27 M31 GCs with 
$M/L_V$ estimates, most of relatively high quality. The median $M/L_V$ is $1.90\pm0.09$ 
with $\sigma=0.46$. The typical uncertainties on $M/L_V$ are 15-20\%.

In Figure 4 we plot $M/L_V$ against dynamical mass. No significant trends are visible,
though the scatter in $M/L_V$ may be higher for lower GC masses.
As a comparison, we have overplotted $M/L_V$ predictions from Maraston (2005) stellar 
population models for three combinations of age and metallicity, assuming a Kroupa IMF. 
The massive object with the very large error bar is 037-327, which has a high and 
uncertain reddening. It has been claimed that 037-327 is the most massive GC in the 
Local Group; assuming that the $\sigma_p$ estimate of Cohen (2006) is correct, 037-327 
is far from the most massive GC---it is, at best, the fifth most massive known in M31.

The three candidate intermediate-age GCs are shown in Figure 4 as large stars. These objects 
have $M/L_V$ entirely consistent with the other objects---there is no compelling evidence that 
they are of intermediate age. The plotted model lines show that the expected difference in 
$M/L_V$ between an old metal-poor GC and a 5 Gyr metal-rich GC is relatively small; below we 
show that the candidate intermediate-age objects are unlikely to be metal-rich.

To our knowledge, there are only two certain intermediate-age star clusters with 
dynamical masses: the 2 Gyr old cluster NGC 1978 in the Large Magellanic Cloud, which 
has a $M/L_V \sim 0.2$ for a single-mass model (Fischer, Welch, \& Mateo 1992), and G114 
in the 3 Gyr merger remnant elliptical NGC 1316, with $M/L_V \sim 1.2$ (Bastian 
\etal~2006). This latter object has a mass of $\sim 1.6 \times 10^{7} M_{\odot}$ and so 
its relevance to the study of typical GCs with lower masses is unclear. Further studies 
of confirmed intermediate-age GCs in the Magellanic Clouds would be valuable to test the 
predictions of stellar population models in this age regime; at least for relatively 
massive clusters, dynamical evolution should not have affected these objects as strongly 
as for old GCs.

\subsection{$M/L_V$ Trends with Metallicity}

D97 reported an inverse correlation between metallicity and $M/L$ in the $K$-band. In the 
$V$ band, their data were consistent with a correlation, but it was only formally 
significant at $1 \sigma$.

Homogeneous spectroscopic metallicities do not exist for all of the GCs studied in this 
paper. However, $V-K$ colors do exist; these are quite sensitive to metallicity, and suffer
only modestly from the age-metallicity degeneracy. We 
derived $V-K$ colors using $V$ mags from Caldwell \etal~(2009) and $K$ mags from the Revised 
Bologna Catalog\footnote{http://www.bo.astro.it/M31/} (see also Galleti \etal~2004). 
Reddening corrections were taken from Barmby \etal~(2000).

Figure 5 shows $M/L_V$ vs.~$(V-K)_0$ for all 27 GCs. There is a strong correlation in the 
same sense found by D97. If we separate the clusters into metal-poor and metal-rich groups 
by dividing at the natural break in the data at $(V-K)_0 = 2.4$ (corresponding to [Fe/H] 
$\sim -1.2$), the median $M/L_V$ of the two populations are $2.11\pm0.13$ and $1.75\pm0.10$. 
A Spearman rank correlation gives $p=0.006$ for a one-tailed test, indicating a significant 
correlation between $M/L_V$ and $(V-K)_0$. If we instead do a linear regression and use the 
Barmby \etal~(2000) relationship between $V-K$ and [Fe/H] derived from Galactic GCs 
(d$(V-K)$/d[Fe/H] = 0.59), we find that d$(M/L)_V$/d[Fe/H] = $-0.38$. If instead we use log 
$M/L_V$, we find a slope of $-0.09$, quite similar to that found by D97 ($-0.1$).

In our view there is no strong evidence from the current dataset that the relation is 
necessarily linear---it could, for example, be a step function. But neither is a linear 
relationship inconsistent with the data. Accurate spectroscopic metallicities will clarify 
this point.

The only object in Figure 5 that is not consistent with the general trends is 405-341, which 
has $M/L_V \sim 1$, half the median value for GCs with $(V-K)_0 < 2.4$. There is nothing 
clearly unusual about the cluster; it has UV-optical colors and a spectrum consistent with 
it being a normal, moderately metal-poor old GC (Rey \etal~2007; Barmby \etal~2000). One 
possibility is that the velocity dispersion of this GC was underestimated by D97, as we 
surmise occurred for 358-219 (discussed in \S 3.4); of course we cannot rule out the cluster 
being a true outlier.

Also plotted in Figure 5 are isochrones for a range of metallicities from Maraston (2005) 
stellar population models. These assume a Kroupa IMF; using a Salpeter IMF instead would 
increase the $M/L_V$ by $\sim 45-57$\% depending on the age and metallicity. The metal-poor 
GCs agree well with predictions for an old stellar population. However, the metal-rich GCs 
have $M/L_V$ lower by about a factor of two than the model predictions for old clusters. In 
\S 5.2 we discuss possible explanations for this difference.

We have also plotted a single 13 Gyr isochrone from Bruzual \& Charlot (2003), assuming a 
Chabrier IMF. The predicted $M/L_V$ is very similar to that of Maraston (2005) for 
metal-poor systems but increases less steeply for metal-rich populations. Nonetheless, the 
basic behavior is the same, and the tension between the models and data remains.

Figure 5 also gives additional information about the candidate intermediate-age GCs discussed in 
\S 3.6. Taken at face value, the model isochrones suggest miminal differences in $M/L_V$ between 
intermediate-age metal-rich GCs and old metal-poor GCs. But because $V-K$ has only a weak 
age-metallicity degeneracy (see the isometallicity contour in Figure 5), the blue $V-K$ colors 
of the candidate intermediate-age GCs are inconsistent with high metallicities for any 
reasonable age. Their properties are as expected for old metal-poor GCs.

\section{The Fundamental Plane}

The fundamental plane of GCs---a tight relationship connecting the velocity dispersion, 
radius, and surface brightness---is analogous to that for elliptical galaxies, as described 
in Djorgovski \& Davis (1987) and Faber \etal~(1987). In ellipticals, the fundamental 
plane reflects the virial theorem plus a ``tilt" corresponding to structural nonhomology and 
stellar population variations. For GCs, the fundamental plane is extremely close to that 
predicted by the virial theorem, indicating only small variations in $M/L$ among GCs 
(Djorgovski 1995). Related arguments for a constant $M/L$ in the cores of Galactic GCs can 
be found in McLaughlin (2000).

The fundamental plane of core parameters ($r_c$, $\mu_0$, and $\sigma_0$) for Galactic GCs from Djorgovski (1995) 
is:

\begin{equation}
\mu_{0} = (-4.9\pm0.2)(\textrm{log } \sigma_0 - 0.45 \textrm{log } r_c) + (20.45\pm0.2)
\end{equation}

Using the revised dynamical and structural data for 38 GCs from McLaughlin \& van der Marel (2005), we refit this 
equation and find:

\begin{equation}
\mu_{0} = (-4.51\pm0.12)(\textrm{log } \sigma_0 - 0.55 \textrm{log } r_c) + (20.17\pm0.10)
\end{equation}

\noindent
where the errors are given by a jackknife. The coefficients agree well within the 
errors. This is not unexpected, since there have been few additional measurements of 
$\sigma_0$ for Galactic GCs in the last decade.

Fitting the core parameters of the M31 GCs alone (excluding 037-327, see below) gives:

\begin{equation}
\mu_{0} = (-4.69\pm0.08)(\textrm{log } \sigma_0 - 0.50 \textrm{log } r_c) + (20.22\pm0.10)
\end{equation}

This is extremely similar to the fundamental plane for Galactic GCs, both in the slope and the 
zeropoint. This agreement shows that there is no significant difference in the $M/L_V$ for GCs in 
M31 and the Galaxy. A small subtlety is that the velocity dispersions for M31 GCs were derived 
from spectroscopy with slit widths comparable to cluster half-light radii, so these ``core" 
parameters are actually tied to the total cluster structure (unlike in the Galaxy, where 
$\sigma_0$ can be more directly measured). So this comparison is not strictly confined to core 
parameters.

Given the good agreement between the M31 and Galactic GCs, we can combine the two samples to get a ``master" 
fundamental plane of 64 GCs. The resulting fit is:

\begin{equation}
\mu_{0} = (-4.68\pm0.06)(\textrm{log } \sigma_0 - 0.52 \textrm{log } r_c) + (20.28\pm0.05)
\end{equation}

\noindent
which gives the best current estimate of the fundamental plane of core parameters for old GCs. Using this equation, in 
Figure 6 we plot the fundamental plane for Galactic and M31 GCs.

We can use this fundamental plane to estimate the $E(B-V)$ for 037-327, the outlier 
among massive clusters in Figure 6. Using the value of $\sigma_0$ in Table 5 and $r_c$ 
from B07, the ``master" plane predicts $\mu_0 = 14.77$. The extinction-corrected value 
from B07, assuming $E(B-V)=1.36$, is $\mu_0 = 13.39$. Agreement is obtained if $E(B-V) 
= 0.92$. This estimate assumes that 037-327 has a similar age to the other M31 GCs used 
in the fundamental plane fit.

\section{Discussion}

\subsection{Intermediate-Age Clusters}

The first result of this paper is that three candidate intermediate-age M31 GCs identified from 
optical spectroscopy have $M/L_V$ and $V-K$ colors identical to old GCs. Thus they are very 
unlikely to be of intermediate-age, assuming standard initial mass functions. A significant hot 
star component---likely to be blue horizontal branch stars---is presumably the cause of the 
strong Balmer lines observed in these clusters. For one of the candidates, 311-033, Rich 
\etal~(2005) present a resolved color-magnitude diagram. It clearly shows a blue horizontal 
branch as decisive proof of an old age.

The conclusion that horizontal branch stars are the cause of the strong Balmer lines in 
these GCs is the same as that reached by Rey \etal~(2007), who present $FUV$ photometry 
for ten candidate intermediate-age M31 GCs. Using a $B-V$ vs.~$FUV-V$ color-color plot, 
they show that nine of these candidates have excess $FUV$ light not expected in younger 
GCs, and thus are probably old GCs with extended blue horizontal branches. The tenth 
object, 126-184, does appear to have colors consistent with an intermediate age, but we 
have shown above that it is also likely to be old.

Together, these observations suggest that a subset of the claimed intermediate-age GCs 
in M31 are old. Additional observations would be needed to extend this conclusion. As 
an example, deep HST imaging has been obtained for several intermediate-age candidates 
from Puzia \etal~(2005), which should be able to determine whether these objects have 
blue horizontal branches (but possibly not blue stragglers, which can also affect GC 
ages derived from spectroscopy; Cenarro \etal~2008). High-resolution spectroscopy of 
more potential intermediate-age GCs would also be fruitful.

\subsection{The Metallicity Trend}

\subsubsection{Comparisons to other Galaxies}

In \S 4 we showed that the $M/L_V$ of metal-poor GCs agree well with model predictions, 
while the metal-rich GCs disagree by $\sim$ a factor of two. It is difficult to assess 
whether this trend is also present in Galactic GCs, since there are few metal-rich clusters 
with dynamical measurements. Figure 19 of McLaughlin \& van der Marel (2005) gives a hint 
of the same behavior for Galactic GCs. This figure plots, for each cluster, the ratio of 
dynamical $M/L$ to that estimated from stellar population models (assuming an age of 13 
Gyr). All six of the GCs with [Fe/H] $> -1$ have ratios lower than the median of the entire 
sample, and the median ratio of these six values is $\sim 0.5$---similar to that inferred 
for the metal-rich GCs in Figure 5. Given the small number of Galactic GCs, this agreement 
is only suggestive, but it does indicate that existing Galactic data are not inconsistent 
with the M31 results.

Neither the Small or Large Magellanic Clouds have any old metal-rich GCs. The cluster R12 in 
M33 is old and moderately metal-rich, and has $M/L_V = 1.03 \pm 0.16$ (Larsen \etal~2002; 
Chandar \etal~2006). NGC 5128, the nearest massive early-type galaxy, now has nearly 30 GCs 
with dynamical mass estimates (Rejkuba \etal~2007; Martini \& Ho 2004). The complication in 
comparing these results to ours is that most of the NGC 5128 GCs studied are very massive ($> 
2-3 \times 10^6 M_{\odot}$), into the mass regime of ``ultra compact dwarfs" or 
``dwarf-globular transition objects" (DGTOs). A number of papers have shown that above this 
mass limit, DGTOs follow a different fundamental plane than GCs; for example, the sizes of 
DGTOs are correlated with their mass, which is not the case for GCs (e.g., Mieske \etal~2006 
and references therein). DGTOs also appear to have high $M/L$, perhaps even higher than 
expected from stellar population models. These observations have been variously attributed to 
the presence of dark matter, tidal effects, multiple stellar populations, or to very long 
relaxation times in these systems (e.g., Dabringhausen \etal~2008).

In any case, we sidestep this issue by considering only those NGC 5128 GCs with colors 
indicating membership in the metal-rich GC subpopulation ($(V-I)_0 > 1.05$) and with masses 
below $2 \times 10^6 M_{\odot}$. In fact, only one object fits these criteria: HGHH92-C41, 
which was observed by both Rejkuba \etal~(2007) and Martini \& Ho (2004). We obtained an 
updated estimate of $M/L_V$ by using the published $\sigma_p$ values from the two studies and 
integrating over the King model parameters given in Rejkuba \etal~(2007). Averaging the two 
values together gives $M/L_V \sim 1.6$. Even though this is a single cluster in NGC 5128, this 
value is indeed consistent with the M31 results.

We conclude that, in the traditional mass range of GCs, all available data are consistent with a notably low 
$M/L_V$ for metal-rich GCs.

\subsubsection{Potential Explanations}

The most straightforward explanation for Figure 5 is that the bulk of the metal-rich GCs in 
M31 are of intermediate-age. However, there is substantial evidence against this 
interpretation. Optical spectroscopy indicates that the majority of the metal-rich GCs are 
old (many of the claimed intermediate-age GCs were classified as having intermediate to low 
metallicities; Beasley \etal~2005, Puzia \etal~2005). Eighteen of the M31 GCs in the 
combined sample of this paper have GALEX $FUV$ photometry. Nearly all have $(FUV-V)_0 \la 
5.5$, inconsistent with expectations for $\sim 5$ Gyr GCs (Rey \etal~2007). Finally, a 
single M31 metal-rich GC (379-312) has an HST/ACS color-magnitude diagram extending 
below the main sequence turnoff; this deep imaging proves an old age for the cluster (Brown 
\etal~2004).

Assuming that most of the GCs are old, natural explanations for the metallicity trend can 
be classified as natal or evolutionary. In the former case, variations in the stellar IMF 
could comfortably explain the differences, if the IMF for metal-rich GCs was either flatter 
or had a higher-mass cutoff than for metal-poor GCs. There is some fragmentary evidence for 
such variations among Galactic GCs. McClure \etal~(1986) used ground-based imaging for a 
small sample of GCs to argue that more metal-poor GCs had steeper main sequence mass 
functions. While this result was challenged by a number of subsequent studies, Djorgovski 
\etal~(1993) used a multivariate statistical analysis of parameters for a large sample of 
GCs to show that metallicity did affect the mass function (though only secondarily---the 
primary parameters are Galactocentric radius and distance above the plane, presumably 
connected to the mass function through dynamical evolution). The mass function 
slope--metallicity relation was interpreted by Smith \& McClure (1987) as evidence for 
self-enrichment, with flatter mass functions resulting in more metal-rich GCs.

While this previous work is not conclusive, the claimed trends go in the right direction 
to explain our result: flatter mass functions in metal-rich GCs will lead to fewer 
low-mass stars for each star of higher mass, and thus to a lower $M/L_V$ than expected 
from a standard initial mass function.

An evolutionary explanation would likely rest on differences in the dynamical state of GCs 
as a function of metallicity. In general, two-body relaxation and the resultant evaporation 
of low-mass stars is expected to lower the $M/L$ (see, e.g., Kruijssen \& Lamers 2008). In 
fact, if one expects all GCs to have lower $M/L$ than predicted by simple stellar 
population models, we could turn the problem around and suggest that metal-rich GCs have 
the ``expected" $M/L$ while that for metal-poor GCs is too large (again, for natal or 
evolutionary reasons). In either case, a metallicity-dependent explanation is needed.

To test for dynamical correlations, we looked for trends between $M/L_V$ and a 
variety of plausible candidate parameters: $r_c$, $r_h$, $c$, the projected 
galactocentric radius, and the relaxation times at the core and half-mass radii. 
These plots are shown in Figure 7. All quantities were taken from B07 or this 
paper. We found no correlations with any of these variables. Formally, 
one-tailed Spearman rank correlation $p$-values ranged from 0.14 to 0.36 (with the 
lowest value for $c$); these can be compared to $p=0.006$ obtained for the $V-K$ 
vs. $M/L_V$ correlation. There is ongoing debate about the extent to which 
present-day structural parameters reflect the dynamical history of Galactic GCs 
(e.g., de Marchi \etal~2007), and most of the GCs in the M31 sample are massive 
and thus probably less evolved than typical Galactic GCs.

More exotic explanations of systematic differences in $M/L$---for example, variations in 
the retention fractions of neutron stars or black holes---do not seem likely to account 
for the large discrepancies, since such objects only account for a few percent of the 
total cluster mass. A caveat is that dynamically evolved GCs may have lost a large 
fraction of their low-mass stars, and perhaps remnants play a larger role in the dynamics 
of the core region, but it is not clear that they can have a large effect on the global 
$M/L$ values. Our result is also unlikely to be related to the increased prevalence of 
luminous low-mass X-ray binary systems in metal-rich GCs (e.g., Kundu \etal~2002), since 
a boost in such systems would be expected to increase the $M/L_V$, not decrease it. 
Another possibility---though difficult for us to constrain---is simply that the stellar 
population model predictions for metal-rich objects are systematically wrong.

In \S 5.2.1, we briefly discussed the high $M/L$ observed for some compact stellar systems 
with masses larger than typical GCs (so-called DGTOs). $M/L$ for DGTOs are frequently 
normalized to the value predicted by stellar populations models so that objects of 
different metallicities can be more easily compared (Mieske \& Kroupa 2008). While 
comparisons of ``normal" old GCs with DGTOs may be fruitful, a central result of this 
paper is that the $M/L_V$ of metal-rich GCs do not match the values predicted by models, and 
such normalization should be done with caution.

Sorting out the differences in $M/L_V$ between metal-rich GCs and DGTOs could have 
extensive implications. If the low $M/L_V$ of metal-rich stellar populations is a general 
phenomenon not confined to GCs, it may be necessarily to revise the estimates of stellar 
masses obtained for early-type galaxies solely from observed luminosities.

\subsection{Distances from the GC Fundamental Plane}

Djorgovski (1995) predicted that with refinement of the GC fundamental plane it could be 
used to estimate distances to galaxies, since of the three parameters ($\sigma$, $\mu$, 
and $r$), only the last was distance dependent.

The $1 \sigma$ error in the $\mu_0$ intercept of the core fundamental plane is 0.05 mag. 
This corresponds to an error of $\sim 5$\% in the distance to the galaxy. Of course, this 
is the error in the basic projection of the plane, and the errors in the data set of 
interest would increase the total error in the distance estimate. In fact, the offset of 
0.05 mag in the intercepts of the Galactic and M31 fundamental planes could be strictly 
interpreted as evidence that the average distance of M31 GCs is $\sim 40$ kpc larger than 
the 780 kpc assumed earlier, but the offset is well within the errors and we make no such 
claim. Another uncertainty is that for M31, the depth of the GC system along the line of 
sight (tens of kpc) is a few percent of the galaxy's distance, though this should not be 
a factor for more distant galaxies.

Counterintuitively, the behavior of GC $M/L_V$ with respect to metallicity actually 
reduces the scatter in the fundamental plane. If GCs exactly followed stellar population 
models, the higher $M/L_V$ of metal-rich GCs compared to metal-poor GCs would increase 
the scatter, since the fundamental plane is essentially the virial theorem combined with 
a $M/L$ term. In the projection of the fundamental plane discussed above, the scatter in 
$\mu$ due to $M/L$ variations will scale as (0.5 log $M/L$) if the virial theorem holds 
exactly. The difference in $M/L_V$ between the metal-poor and metal-rich GCs is less than 
20\% in the median, corresponding to $\sim 0.04$ mag in $\mu$---a very small effect.

We suggest that the spread in the fundamental plane is small enough that it can be used 
as an additional distance constraint for nearby galaxies. Another potential use is to 
estimate reddening for individual GCs in a given galaxy, as we have done for 037-327.

\acknowledgments

We thank an anonymous referee for a critical reading of the manuscript. J.~S.~was supported by 
NASA through a Hubble Fellowship, administered by the Space Telescope Science Institute, 
which 
is operated by the Association of Universities for Research in Astronomy, Incorporated, under 
NASA contract NAS5-26555. J.~B.~acknowledges support from the NSF through grant AST-0808099. 
J.~P.~H. was supported by the Smithsonian Institution. S.~L.~acknowledges support by NWO VIDI grant 639.042.610.
J.S.~is indebted to Dave Latham for his 
hospitality when the paper was being finished. The data presented herein were obtained at the 
W.~M.~Keck Observatory, which is operated as a scientific partnership among the California 
Institute of Technology, the University of California, and the National Aeronautics and Space 
Administration. The Observatory was made possible by the generous financial support of the 
W.~M.~Keck Foundation. We acknowledge Steve Vogt and the entire HIRES team for building an 
instrument with such excellent long-term stability.

\newpage

\begin{figure}
\epsscale{0.85}
\plotone{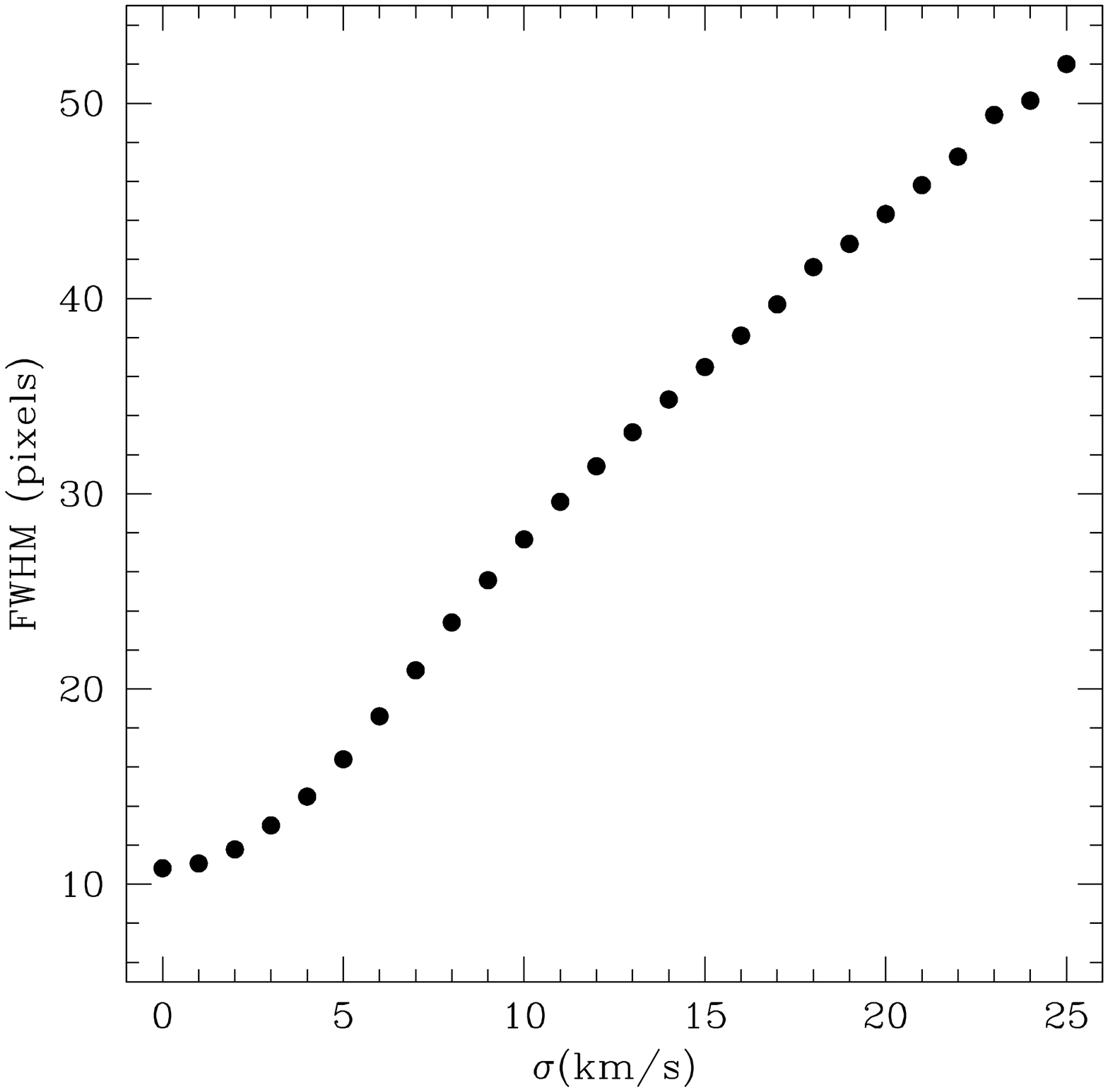}
\figcaption[fig1.eps]{\label{fig:fig1}
Sample relationship between the full-width at half-maximum (FWHM) of the cross-correlation function
between template stars and the input velocity dispersion ($\sigma$) of the convolving Gaussian.}
\end{figure}

\begin{figure}    
\epsscale{0.85}   
\plotone{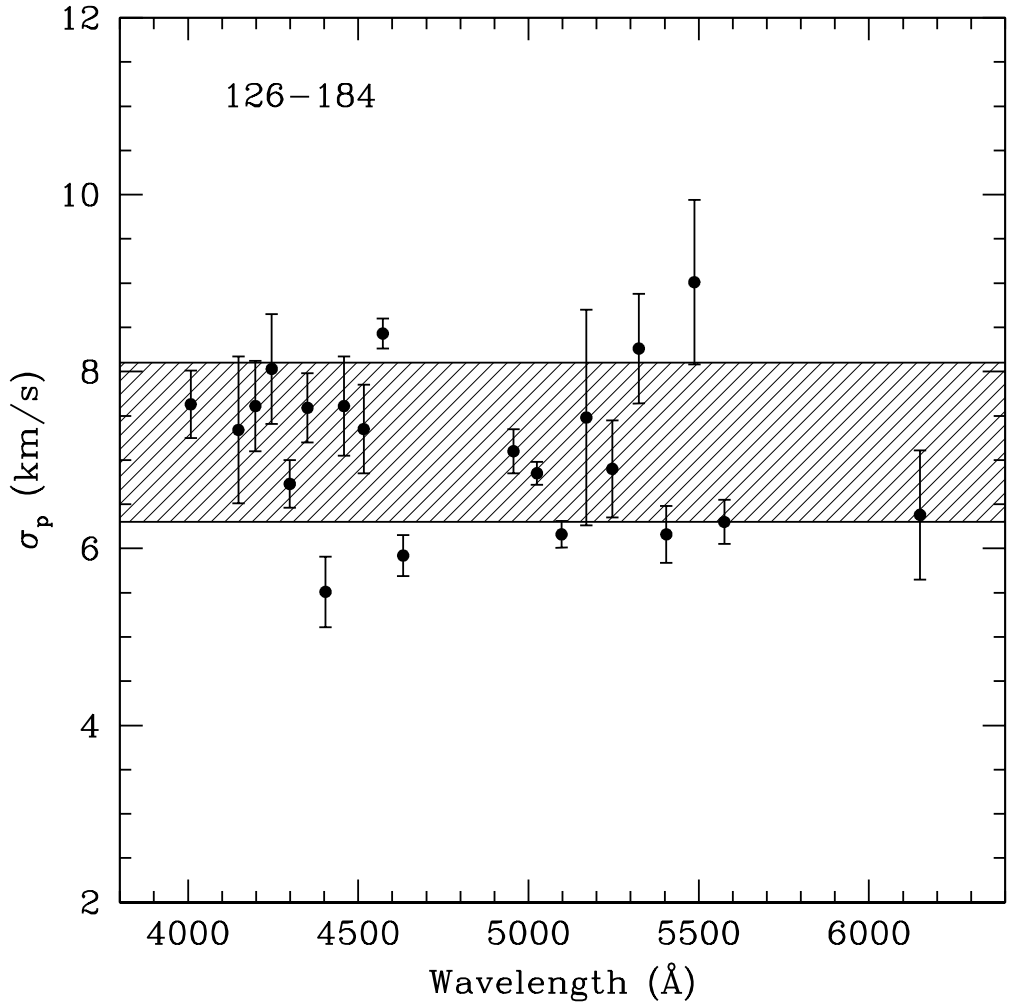}
\figcaption[fig2.eps]{\label{fig:fig2}
Order-by-order values of $\sigma_p$ for 126-184, with the rms variation among different template pairs plotted 
as error bars. The shaded region represents the 1$\sigma$ boundary of the overall $\sigma_p$ estimate. The error 
in $\sigma_p$ appears to be dominated by variation among orders rather than by variation among templates.}
\end{figure}

\begin{figure}  
\epsscale{0.85} 
\plotone{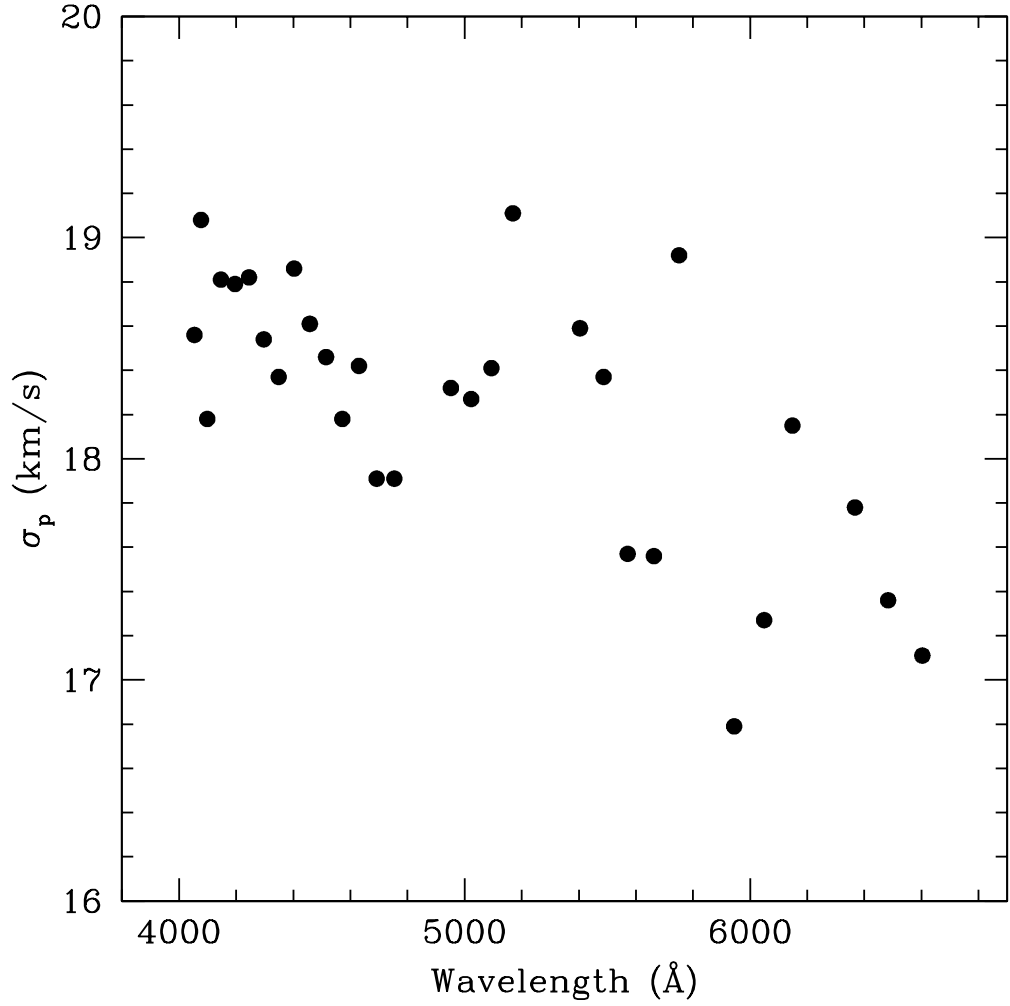}
\figcaption[fig3.eps]{\label{fig:fig3}
Projected velocity dispersion $\sigma_p$ as a function of wavelength for Keck/HIRES data of 163-217.
The data show a quasi-monotonic decrease in $\sigma_p$ toward redder wavelengths.}
\end{figure}

\begin{figure}
\epsscale{0.85}
\plotone{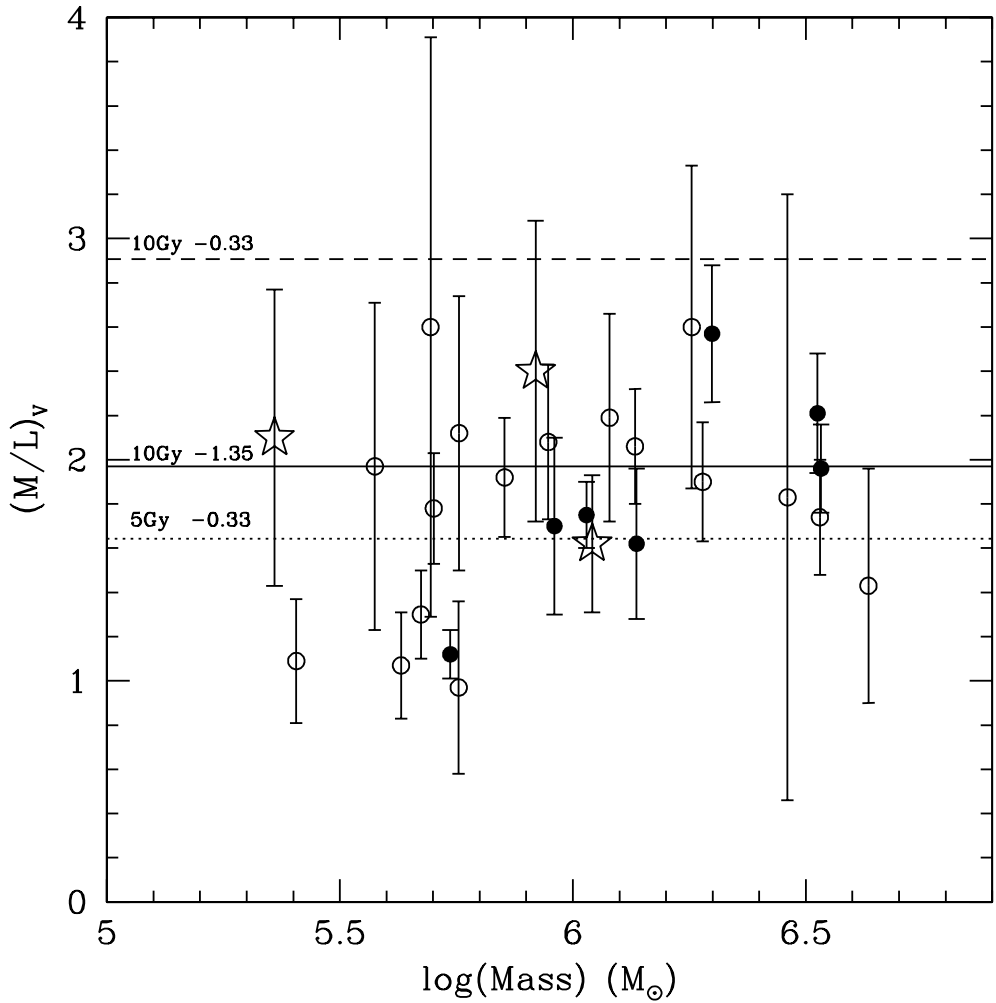}
\figcaption[fig4.eps]{\label{fig:fig4}
$M/L_V$ of M31 GCs. Data from the literature are open circles, data from this paper for old GCs are 
filled circles, and candidate intermediate-age GCs are large stars. Maraston (2005) stellar population model 
predictions are plotted for three representative combinations: 10 Gyr, [$Z$/H] = $-1.35$ (solid line); 10 
Gyr, [$Z$/H] = $-0.33$ (dashed line); 5 Gyr, [$Z$/H] = $-0.33$ (dotted line). The objects are all 
consistent with an essentially constant $M/L_V$ with mass, and the intermediate-age candidates have 
$M/L_V$ consistent with the bulk of the other objects---they are probably old GCs.}
\end{figure}

\begin{figure}
\epsscale{0.85}
\plotone{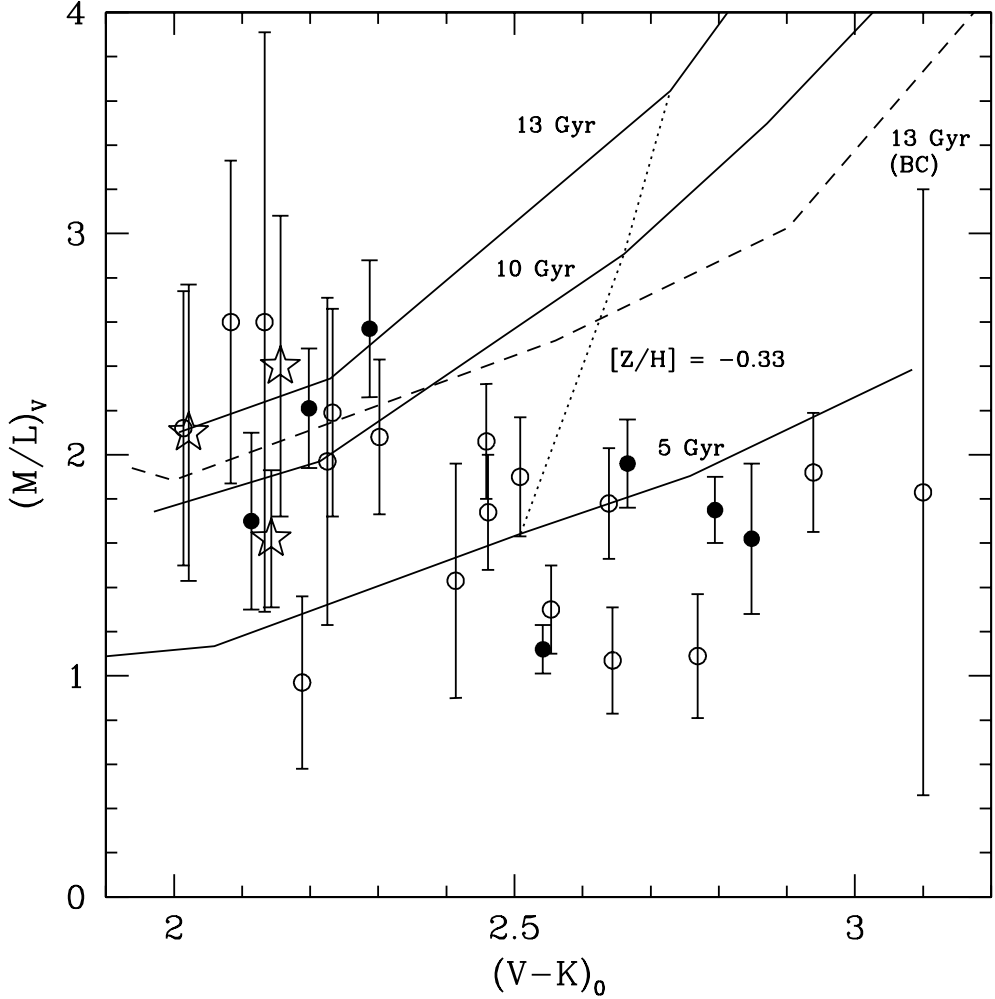}
\figcaption[fig5.eps]{\label{fig:fig5}
$M/L_V$ vs.~$(V-K)_0$ for M31 GCs. Symbols as in Figure 4. $M/L_V$ of the metal-rich GCs is 
systematically lower than for metal-poor GCs. Maraston (2005) stellar population models
assuming a Kroupa IMF are plotted for three ages: 5, 10, and 13 Gyr (solid lines).
A 13 Gyr Bruzual \& Charlot (2003) model assuming a Chabrier IMF is also plotted (dashed line).
The dotted line is an isometallicity contour for the Maraston models, indicating the weak age-metallicity
degeneracy in $V-K$. The $M/L_V$ trend with metallicity is inconsistent with naive expectations from 
stellar population models.}
\end{figure}

\begin{figure}
\epsscale{0.85}
\plotone{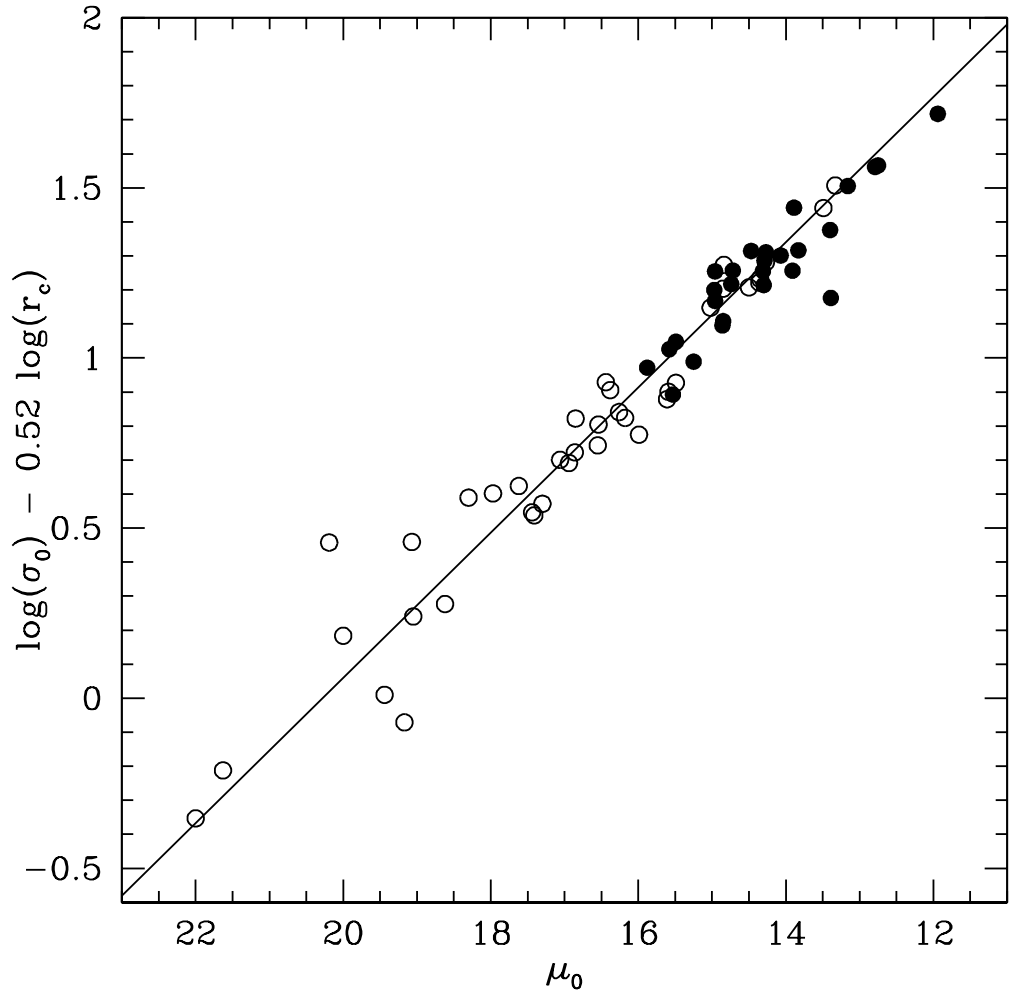}
\figcaption[fig6.eps]{\label{fig:fig6}
Fundamental plane of core parameters for Galactic (open circles) and M31 (filled circles) GCs. Objects from the two
galaxies follow essentially the same fundamental plane, indicating very similar mass-to-light ratios.}
\end{figure}

\begin{figure}
\epsscale{0.85}
\plotone{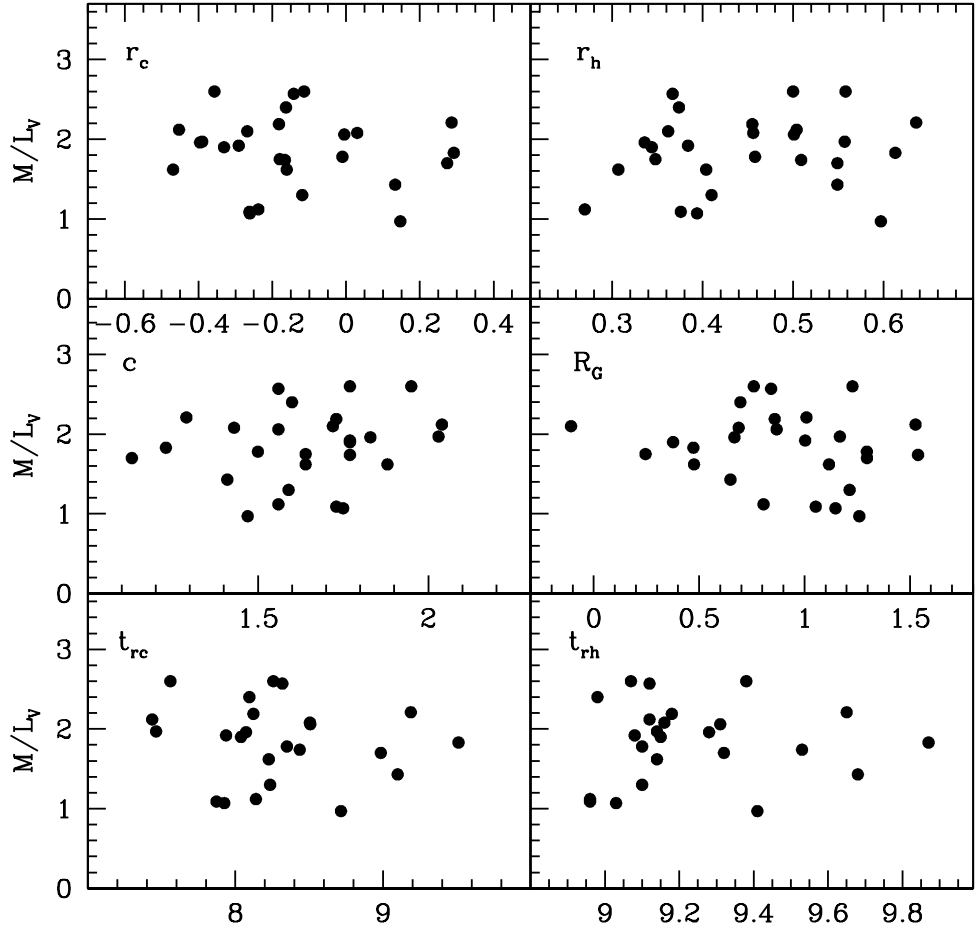}
\figcaption[fig7.eps]{\label{fig:fig7}
$M/L_V$ plotted against six dynamically relevant quantities for M31 GCs: core and half-light radii (in pc), concentration,
galactocentric radius ($R_{G}$; in kpc), and the core and half-light relaxation times ($t_{rc}$, $t_{rh}$; in yr). All abscissae 
are plotted in the logarithm. Values are taken from Barmby \etal~(2007) or this paper. No significant correlations are 
apparent.}
\end{figure}

\begin{deluxetable}{lccc}
\tablewidth{0pt}
\tablecaption{Observing Log
        \label{tab:obslog}}
\tablehead{ID & Slit & Exp. Time & Date  \\
           &  & (hr) &  }
 
\startdata

034-096 & $0.87\arcsec \times 5\arcsec$ & 1.0 & Oct 1997 \\ 
163-217 & $0.87\arcsec \times 5\arcsec$ & 1.5 & Oct 1997 \\
171-222 & $0.87\arcsec \times 5\arcsec$ & 1.5 & Oct 1997 \\
225-280 & $0.87\arcsec \times 5\arcsec$ & 1.5 & Oct 1997 \\
\hline
126-184 & $0.86\arcsec \times 7\arcsec$    & 0.5 & Oct 2005 \\
232-286 & $0.86\arcsec \times 7\arcsec$    & 0.5 & Oct 2005 \\
311-033 & $0.86\arcsec \times 7\arcsec$    & 0.5 & Oct 2005 \\
358-219 & $0.86\arcsec \times 7\arcsec$    & 0.5 & Oct 2005 \\
U77 (M33) & $0.86\arcsec \times 7\arcsec$  & 0.5 & Oct 2005 \\
\hline
338-076 & $0.86\arcsec \times 7\arcsec$ & 3.0 & Oct 2006 \\
\hline
006-058 & $0.86\arcsec \times 7\arcsec$ & 2.5 & Sep/Oct 2007 \\
058-119 & $0.86\arcsec \times 7\arcsec$ & 4.5 & Sep/Oct 2007 \\
163-217 & $0.86\arcsec \times 7\arcsec$ & 3.5 & Sep/Oct 2007 \\
171-222 & $0.86\arcsec \times 7\arcsec$ & 2.5 & Sep/Oct 2007 \\
225-280 & $0.86\arcsec \times 7\arcsec$ & 3.5 & Sep/Oct 2007 \\
358-219 & $0.86\arcsec \times 7\arcsec$ & 4.0 & Sep/Oct 2007 \\

\enddata
\end{deluxetable}

\begin{deluxetable}{lcc}
\tablewidth{0pt}
\tablecaption{New Observed Velocity Dispersions and Radial Velocities
        \label{tab:sigobs}}
\tablehead{ID/year & $\sigma_p$ & $v_{helio}$ \\
           &  (km/s) & (km/s) }

\startdata
006-058        & $12.1\pm0.3$ & $-234.3\pm1.6$ \\
034-096        & $12.8\pm0.6$ & \nodata \\
058-119        & $21.0\pm0.9$ & $-222.9\pm1.7$ \\
126-184        & $7.2\pm0.9$  & $-166.2\pm1.6$ \\
163-217 (1997) & $18.2\pm1.0$ & \nodata \\
163-217 (2007) & $18.3\pm0.6$ & $-162.1\pm1.8$ \\
171-222 (1997) & $15.0\pm0.5$ & \nodata \\
171-222 (2007) & $15.0\pm0.5$ & $-264.6\pm1.6$ \\
225-280 (1997) & $28.4\pm1.3$ & \nodata \\
225-280 (2007) & $28.5\pm0.8$ & $-159.1\pm1.9$ \\
232-286        & $13.4\pm1.8$ & $-186.3\pm1.9$ \\
311-033        & $14.9\pm1.3$ & $-514.8\pm1.8$ \\
338-076        & $20.7\pm0.9$ & $-260.4\pm1.8$ \\
358-219 (2005) & $11.8\pm1.3$ & $-313.5\pm2.6$ \\
358-219 (2007) & $12.1\pm1.3$ & $-309.2\pm1.7$ \\
U77-M33        & $4.6\pm1.0$  & $-218.2\pm2.9$ \\
\enddata
\end{deluxetable}

\begin{deluxetable}{lcccccccc}
\tablewidth{0pt}
\rotate
\tabletypesize{\scriptsize}
\tablecaption{New Dynamical Quantities for M31 GCs
        \label{tab:dynus}}
\tablehead{ID & $\sigma_0$ & $\sigma_{\infty}$  & $M_{King}$ & $M_{vir}$ & $E_b$ & $L_{tot}$ & $(M_{king}/L)_V$ & $(M_{vir}/L)_V$  \\
           &  (km/s) & (km/s) & ($M_{\odot}$) & ($M_{\odot}$) & (erg) & ($L_{\odot}$) & & }

\startdata
006-058 & $13.6\pm0.3$ & $11.0\pm0.3$ & $(5.18\pm0.26) \times 10^5$ & $(5.46\pm0.27) \times 10^5$ & $3.5 \times 10^{51}$ & $(4.90\pm0.34) \times 10^5$ & $1.06\pm0.10$ & $1.12\pm0.11$ \\
058-119 & $23.3\pm1.0$ & $18.9\pm0.8$ & $(1.89\pm0.16) \times 10^6$ & $(1.99\pm0.17) \times 10^6$ & $3.8 \times 10^{52}$ & $(7.76\pm0.53) \times 10^5$ & $2.43\pm0.29$ & $2.57\pm0.31$ \\
163-217 & $21.0\pm0.7$ & $16.2\pm0.5$ & $(1.16\pm0.08) \times 10^6$ & $(1.37\pm0.09) \times 10^6$ & $1.8 \times 10^{52}$ & $(8.45\pm1.70)^{\ast} \times 10^5$ & $1.37\pm0.29$ & $1.62\pm0.34$ \\
171-222 & $16.7\pm0.6$ & $13.4\pm0.4$ & $(9.83\pm0.66) \times 10^5$ & $(1.07\pm0.07) \times 10^6$ & $1.0 \times 10^{52}$ & $(6.10\pm0.35)^{\ast} \times 10^5$ & $1.61\pm0.14$ & $1.75\pm0.15$ \\
225-280 & $32.4\pm0.9$ & $25.2\pm0.7$ & $(2.99\pm0.17) \times 10^6$ & $(3.41\pm0.19) \times 10^6$ & $1.1 \times 10^{53}$ & $(1.74\pm0.12) \times 10^6$ & $1.72\pm0.17$ & $1.96\pm0.20$ \\
338-076 & $22.3\pm1.0$ & $18.5\pm0.8$ & $(3.26\pm0.28) \times 10^6$ & $(3.35\pm0.29) \times 10^6$ & $5.9 \times 10^{52}$ & $(1.51\pm0.10) \times 10^6$ & $2.15\pm0.26$ & $2.21\pm0.27$ \\
358-219 & $13.0\pm1.4$ & $10.9\pm1.2$ & $(8.87\pm1.92) \times 10^5$ & $(9.13\pm1.98) \times 10^5$ & $5.2 \times 10^{51}$ & $(5.37\pm0.37) \times 10^5$ & $1.65\pm0.38$ & $1.70\pm0.40$ \\
\hline
126-184 & $8.1 \pm1.0$ & $6.4 \pm0.8$ & $(2.11\pm0.53) \times 10^5$ & $(2.29\pm0.57) \times 10^5$ & $5.0 \times 10^{50}$ & $(1.09\pm0.22)^{\ast} \times 10^5$ & $1.93\pm0.62$ & $2.10\pm0.67$ \\
232-286 & $14.9\pm2.0$ & $12.0\pm1.6$ & $(7.80\pm2.09) \times 10^5$ & $(8.32\pm2.24) \times 10^5$ & $6.4 \times 10^{51}$ & $(3.47\pm0.24) \times 10^5$ & $2.25\pm0.63$ & $2.40\pm0.68$ \\
311-033 & $16.5\pm1.4$ & $13.2\pm1.2$ & $(1.02\pm0.18) \times 10^6$ & $(1.10\pm0.19) \times 10^6$ & $1.0 \times 10^{52}$ & $(6.76\pm0.46) \times 10^5$ & $1.50\pm0.29$ & $1.62\pm0.31$ \\
\enddata
\tablecomments{Luminosities marked with an asterisk are derived in this paper (see text); all others are from Barmby \etal~(2007). The 
three clusters in the second part of the table are the candidate intermediate-age objects.}
\end{deluxetable}

\begin{deluxetable}{lccc}
\tablewidth{0pt}
\tablecaption{Comparisons of Central Velocity Dispersions with Published Values
        \label{tab:sigcomp2}}
\tablehead{ID & $\sigma_0$ (us) & $\sigma_0$ (D97) & $\sigma_0$ (DG97)  \\
           &  (km/s) & (km/s) & (km/s) }

\startdata
006-058 & $13.6\pm0.3$ & $13.2\pm0.7$  & $12.5\pm0.5$ \\
225-280 & $32.4\pm0.9$ & $30.0\pm2.3$  & $32.2\pm0.6$ \\
358-219 & $13.0\pm1.0$ & $8.9\pm1.9$   & $7.8\pm2.0$ \\
\enddata
\end{deluxetable}

\begin{deluxetable}{lccccccccr}
\tablewidth{0pt}
\rotate
\tabletypesize{\scriptsize}
\tablecaption{Rederived Dynamical Quantities for M31 GCs 
        \label{tab:dynlit}}
\tablehead{ID & $\sigma_0$ & $\sigma_{\infty}$  & $M_{King}$ & $M_{vir}$ & $E_b$ & $L_{tot}$ & $(M_{king}/L)_V$ & $(M_{vir}/L)_V$ & source  \\
           &  (km/s) & (km/s) & ($M_{\odot}$) & ($M_{\odot}$) & (erg) & ($L_{\odot}$) & & &  }

\startdata

000-001	& $26.2\pm1.8$ & $20.6\pm1.4$ & $(3.0\pm0.4) \times 10^{6}$	&	$(3.4\pm0.5) \times 10^{6}$	&	$7.7 \times 10^{52}$	& $(1.95\pm0.04) \times 10^{6}$	& $1.56\pm0.23$	& $1.74\pm0.26$ &	D97;C06  \\
000-002	& $11.2\pm1.6$ & $8.4\pm1.2$  & $(4.5\pm1.3) \times 10^{5}$	&	$(5.7\pm1.7) \times 10^{5}$	&	$1.9 \times 10^{51}$	& $(2.69\pm0.06) \times 10^{5}$	& $1.69\pm0.50$	& $2.12\pm0.62$ &	D97		\\
012-064	& $18.0\pm2.4$ & $14.1\pm1.9$ & $(1.6\pm0.4) \times 10^{6}$	&	$(1.8\pm0.5) \times 10^{6}$	&	$1.9 \times 10^{52}$	& $(6.92\pm0.33) \times 10^{5}$	& $2.33\pm0.65$	& $2.60\pm0.73$ &	D97  \\
020-073	& $16.4\pm0.9$ & $13.3\pm0.8$ & $(1.3\pm0.2) \times 10^{6}$	&	$(1.4\pm0.2) \times 10^{6}$	&	$1.3 \times 10^{52}$	& $(6.61\pm0.15) \times 10^{5}$	& $1.95\pm0.25$	& $2.06\pm0.26$ &	D97;DG97 \\
023-078	& $27.9\pm5.1$ & $23.0\pm4.2$ & $(4.2\pm1.6) \times 10^{6}$	&	$(4.3\pm1.6) \times 10^{6}$	&	$1.2 \times 10^{53}$	& $(3.02\pm0.14) \times 10^{6}$	& $1.39\pm0.52$	& $1.43\pm0.53$ &	D97 \\
037-327	& $21.3\pm3.2$ & $17.8\pm2.6$ & $(2.8\pm0.8) \times 10^{6}$	&	$(2.9\pm0.9) \times 10^{6}$	&	$4.6 \times 10^{52}$	& $(1.57\pm1.17) \times 10^{6}$\tablenotemark{a}& $1.79\pm1.34$	& $1.83\pm1.37$ &	 C06 \\
045-108	& $10.1\pm0.8$ & $7.9\pm0.6$  & $(7.8\pm1.2) \times 10^{5}$	&	$(8.9\pm1.4) \times 10^{5}$	&	$2.8 \times 10^{51}$	& $(4.27\pm0.20) \times 10^{5}$	& $1.82\pm0.31$	& $2.08\pm0.35$ &	D97;DG97 \\
158-213	& $24.6\pm1.6$ & $19.5\pm1.2$ & $(1.7\pm0.2) \times 10^{6}$	&	$(1.9\pm0.2) \times 10^{6}$	&	$3.8 \times 10^{52}$	& $(1.00\pm0.05) \times 10^{6}$	& $1.73\pm0.25$	& $1.90\pm0.27$ &	D97;DG97  \\
240-302	& $13.2\pm1.3$ & $10.1\pm1.0$ & $(1.0\pm0.2) \times 10^{6}$	&	$(1.2\pm0.2) \times 10^{6}$	&	$6.0 \times 10^{51}$	& $(5.50\pm0.26) \times 10^{5}$	& $1.81\pm0.39$	& $2.19\pm0.47$ &	D97 \\
289-000	& $9.6\pm2.4$  & $7.2\pm1.8$  & $(4.0\pm2.0) \times 10^{5}$	&	$(5.0\pm2.5) \times 10^{5}$	&	$1.2 \times 10^{51}$	& $(1.91\pm0.04) \times 10^{5}$	& $2.12\pm1.07$	& $2.60\pm1.31$ &	D97  \\
343-105	& $11.3\pm2.1$ & $8.9\pm1.6$  & $(3.4\pm1.3) \times 10^{5}$	&	$(3.8\pm1.4) \times 10^{5}$	&	$1.6 \times 10^{51}$	& $(1.91\pm0.09) \times 10^{5}$	& $1.77\pm0.67$	& $1.97\pm0.74$ &	D97;DG97 \\
373-305	& $14.4\pm0.9$ & $11.4\pm0.7$ & $(6.5\pm0.8) \times 10^{5}$	&	$(7.1\pm0.9) \times 10^{5}$	&   $4.9 \times 10^{51}$	& $(3.72\pm0.09) \times 10^{5}$	& $1.75\pm0.24$	& $1.92\pm0.27$ &	D97 \\
379-312	& $9.4\pm1.1$  & $7.6\pm0.9$  & $(2.4\pm0.6) \times 10^{5}$	&	$(2.6\pm0.6) \times 10^{5}$	&    $7.9 \times 10^{50}$	& $(2.34\pm0.11) \times 10^{5}$	& $1.03\pm0.26$	& $1.09\pm0.28$ &	D97 \\
384-319	& $10.8\pm0.7$ & $8.7\pm0.6$  & $(4.5\pm0.6) \times 10^{5}$	&	$(4.7\pm0.6) \times 10^{5}$	&	$2.0 \times 10^{51}$	& $(3.63\pm0.17) \times 10^{5}$	& $1.23\pm0.19$	& $1.30\pm0.20$ &	D97;DG97 \\
386-322	& $13.2\pm1.4$ & $10.8\pm1.1$ & $(4.1\pm0.9) \times 10^{5}$	&	$(4.3\pm0.9) \times 10^{5}$	&	$2.7 \times 10^{51}$	& $(3.98\pm0.19) \times 10^{5}$	& $1.03\pm0.23$	& $1.07\pm0.24$ &	D97 \\
405-351	& $9.3\pm1.8$  & $7.6\pm1.5$  & $(5.5\pm2.2) \times 10^{5}$	&	$(5.7\pm2.3) \times 10^{5}$	&	$1.8 \times 10^{51}$	& $(5.89\pm0.14) \times 10^{5}$	& $0.93\pm0.37$	& $0.97\pm0.39$ &	D97 \\
407-352	& $10.5\pm0.7$ & $8.6\pm0.5$  & $(4.8\pm0.6) \times 10^{5}$	&	$(5.0\pm0.6) \times 10^{5}$	&	$2.0 \times 10^{51}$	& $(2.82\pm0.07) \times 10^{5}$	& $1.72\pm0.24$	& $1.78\pm0.25$ &	D97  \\

\enddata
\tablenotetext{a}{This luminosity is calculated by assuming $E(B-V)=0.92$ rather than the $E(B-V)=1.36$ used by Barmby \etal~2007.} 
\end{deluxetable}

\begin{deluxetable}{lccccl}
\tablewidth{0pt}
\rotate
\tabletypesize{\footnotesize}
\tablecaption{New Structural Parameters for M31 GCs
        \label{tab:struct}}
\tablehead{ID & $r_c$ & $c$ & $r_h$ & $\mu_0$ & Filter   \\
           &  (pc) &  & (pc) & (mag) }

\startdata

126-184 & $0.54\pm0.05$ & $52.7\pm1.9$ & $2.30\pm0.21$ & $15.97\pm0.03$ & WFPC2/F555W \\
163-217 & $0.34\pm0.03$ & $75.4\pm2.7$ & $2.03\pm0.07$ & $13.33\pm0.02$ & WFPC2/F555W \\
171-222 & $0.66\pm0.06$ & $43.2\pm1.0$ & $2.23\pm0.09$ & $15.34\pm0.03$ & ACS/F435W \\

\enddata
\end{deluxetable}

\end{document}